\documentclass{jfm}

\usepackage{graphicx}
\usepackage{natbib}
\usepackage[labelfont=it]{subcaption}

\ifCUPmtlplainloaded \else
  \checkfont{eurm10}
  \iffontfound
    \IfFileExists{upmath.sty}
      {\typeout{^^JFound AMS Euler Roman fonts on the system,
                   using the 'upmath' package.^^J}%
       \usepackage{upmath}}
      {\typeout{^^JFound AMS Euler Roman fonts on the system, but you
                   dont seem to have the}%
       \typeout{'upmath' package installed. JFM.cls can take advantage
                 of these fonts,^^Jif you use 'upmath' package.^^J}%
      }
  \else
  \fi
\fi

\ifCUPmtlplainloaded \else
  \checkfont{msam10}
  \iffontfound
    \IfFileExists{amssymb.sty}
      {\typeout{^^JFound AMS Symbol fonts on the system, using the
                'amssymb' package.^^J}%
       \usepackage{amssymb}%
       \let\le=\leqslant  
         
      }{}
  \fi
\fi

\ifCUPmtlplainloaded \else
  \IfFileExists{amsbsy.sty}
    {\typeout{^^JFound the 'amsbsy' package on the system, using it.^^J}%
     \usepackage{amsbsy}}
    {\providecommand\boldsymbol[1]{\mbox{\boldmath $##1$}}}
\fi

\providecommand\bnabla{\boldsymbol{\nabla}}
\providecommand\bcdot{\boldsymbol{\cdot}}

\newsavebox{\astrutbox}
\sbox{\astrutbox}{\rule[-5pt]{0pt}{20pt}}

\newcommand\etal{\mbox{\textit{et al.}}}

\newcommand{\rd}{\mathrm{d}}
\usepackage{color,bm,amsmath,hyperref}

\title[Flow rate--pressure drop relation for deformable microchannels]{Flow rate--pressure drop relation for deformable shallow microfluidic channels}

\author[I. C. Christov, V. Cognet, T. C. Shidhore and H. A. Stone]%
{Ivan C. Christov$^{1,3}$%
  \thanks{Email addresses for correspondence: christov@purdue.edu, hastone@princeton.edu},\ns
Vincent Cognet$^{1,2}$,\ns Tanmay C. Shidhore$^{3}$\ns and Howard A. Stone$^{1}\dagger$}
\affiliation{$^1$Department of Mechanical and Aerospace Engineering, Princeton University,\\ Princeton, New Jersey 08544, USA\\[\affilskip]
$^2$\'Ecole Normale Sup\'erieure de Cachan, Cachan Cedex, France\\[\affilskip]
$^3$School of Mechanical Engineering, Purdue University, West Lafayette, Indiana 47907, USA
}

\pubyear{???}
\volume{???}
\pagerange{??--??}
\date{?; revised ?; accepted ?. Today is \today}
\begin{document}

\maketitle

\begin{abstract}
Laminar flow in devices fabricated from soft materials causes deformation of the passage geometry, which affects the flow rate--pressure drop relation. For a given pressure drop, in channels with narrow rectangular cross-section, the flow rate varies as the cube of the channel height, so deformation can produce significant quantitative effects, including nonlinear dependence on the pressure drop [{Gervais, T., El-Ali, J., G\"unther, A. \& Jensen, K.\ F.}\ 2006 Flow-induced deformation of shallow microfluidic channels.\ \textit{Lab Chip} \textbf{6}, 500--507]. Gervais \etal\ proposed a successful model of the deformation-induced change in the flow rate by heuristically coupling a Hookean elastic response with the lubrication approximation for Stokes flow. However, their model contains a fitting parameter that must be found for each channel shape by performing an experiment. We present a perturbation approach for the flow rate--pressure drop relation in a shallow deformable microchannel using the theory of isotropic quasi-static plate bending and the Stokes equations under a lubrication approximation (specifically, the ratio of the channel's height to its width and of the channel's height to its length are both assumed small). Our result contains no free parameters and confirms Gervais \etal's observation that the flow rate is a quartic polynomial of the pressure drop. The derived flow rate--pressure drop relation compares favorably with experimental measurements.
\end{abstract}

\section{Introduction}

Fluid--structure interactions in low-Reynolds-number flows occur at many scales and across various physical contexts \citep{BGN14,DS16}. For example, microfluidic channels created using polymers such as poly(dimethylsiloxane) (PDMS) are soft \citep{LOVB97,XW98}. As a result, deformation away from their rectangular cross-sectional molding should be expected, and indeed has been observed under typical flow conditions \citep{HKBC03,GEGJ06,SLHLUB09,HUZK09,CTS12,RS16,RDC17}. It is well known that, in a laminar viscous flow through a channel of arbitrary but fixed cross-section specified \textit{a priori}, the pressure drop $\Delta p$ across the channel is proportional to the volumetric flow rate $q$ through it \citep{HB83,S93,B08}. Moreover, the proportionality constant, which is termed the \emph{hydrodynamic resistance}, is only a function of the fluid's viscosity and the geometry of the channel \citep{HB83,B08}. While such a $q$--$\Delta p$ relation is valid for sufficiently low flow rates and low pressure drops, a markedly nonlinear regime due to fluid--structure interactions has been observed experimentally at higher values of $q$ or $\Delta p$ \citep{GEGJ06,SLHLUB09,HUZK09,CTS12}. \citet{GEGJ06} proposed a simple model (with one fitting parameter) of the flow rate--pressure drop relationship in a soft shallow microchannel. Based on the assumption of a Hookean response of the elastic channel walls to the fluid pressure, \citet{GEGJ06} related the channel shape to the hydrodynamic pressure, and inserted this relation into the standard relationship from lubrication theory between the pressure gradient and the flow rate. While this approach was successful in describing experimental measurements, it is not a complete theory.  In the present work, we use asymptotic analysis to find the flow rate--pressure drop relation in the bending-dominated regime of channel deformation, without any fitting parameters.

Obtaining analytical expressions, even approximate ones, for such a ``generalized Poi\-seuille's law'' is of importance for the design and construction of microfluidic devices \citep{S11,SMMC11,OYE13,AS15,RS16,RDC17,GMV17}, where the compliance of soft polymeric materials allows higher throughput than their rigid counterparts. In turn, this degree of freedom makes it possible to manufacture various valves, pumps and self-regulating fluid control elements for lab-on-a-chip technologies \cite[see, \textit{e.g.},][Section III.C]{SQ05}. Similarly, on the basis of the flow--deformation coupling, \citet{HKBC03} proposed using deformable microchannels as microfluidic diffusion diluters. Beyond microfluidics, a flow rate--pressure drop relation is also necessary for, \textit{e.g.}, upscaling deformable porous media \citep{IMP08}, modeling multiphase fluid--structure interactions in industrial piping and turbomachinery \citep{ASIDSCB11}, and understanding biofluid mechanics of blood vessels \citep{P80,GJ04}.

In fact, similar types of fluid--structure interaction problems  have a history in biofluid mechanics, before the modern advent of microfluidics. For example, Fung, in his {\it Biomechanics} textbook \cite[][\S3.4]{F97}, derived a flow rate--pressure drop relation for steady, laminar flow in an elastic tube starting from Poiseuille's relation for a pipe of uniform radius and substituting an axially-varying radius, which is found as a solution to Hooke's law accounting for only circumferential strains. An important consequence of this result is that $q$ is a \emph{nonlinear} function of $\Delta p$, which has significant implications for biofluid mechanics of flow through soft tubes \citep{RK72,GJ04}, specifically in the study of collapsible blood vessels and constrictions \citep{C69,KCM69,P80,KHG12}.  In this context, a perturbative approach to the derivation of the related ``tube laws'' (relationships between pressure drop and cross-sectional area of a tube) from shell theory has proven fruitful \citep{WHJW10}. 

Here, we are interested in the steady-state response of a a microchannel due to flow through it. Nevertheless, the transient deformation problem for a microchannel has been studied experimentally \citep{DGPHD07,PYDHD09}, showing that the characteristic response time can be found in terms of the system's geometric parameters, the fluid's viscosity and the elastic solid's Young's modulus. A lubrication model captures the bulk of the transient response observed in experiments \citep{PYDHD09}, thereby verifying the scaling analysis of \cite{DGPHD07}; \citet{MCC13} extended the  one-dimensional lubrication model of \citet{PYDHD09} to account for electroosmotic flows. It is of interest to account for the effect of fluid--structure interactions in such flows because, for example, flow-wise cross-sectional variations affect electrokinetics and increase electrolyte dispersion  \citep{G02,BBS12} and can be used to improve the sensitivity of impedance-based flow rate measurements \citep{NNBR17}. More recently, \citet{EG14,EG16} analyzed, using perturbation methods, several problems of axisymmetric axial viscous flows in soft cylinders, obtaining closed-form leading-order solutions of the transient response of the elastic shell and  the corresponding time evolution of the fluid pressure.

Beyond one-dimensional models, fewer works have considered the full three-dimensional response of the microchannel. \citet{GEGJ06} performed steady-state fluid--structure interaction simulations, while  \citet{OYE13} simulated only the fluid flow in the deformed channel by using their experimentally measured wall deformation to set the geometry. Meanwhile, \citet{CPFY12} employed a three-dimensional computational model, solving numerically the coupled equations of fluid mechanics and elasticity. However, \citet{CPFY12} showed that a two-dimensional model based on assuming the elastic wall is an infinitely wide elastic beam of finite thickness is a reasonable approximation to the three-dimensional deformations observed in experiments. Thus, it has been established that there is value in lower-dimensional models, especially if the models can be analyzed completely.

To this end, in this paper, we begin by deriving the governing equations of lubrication theory  for long, shallow microchannels in \S\ref{sec:lubri}, via a leading-order asymptotic solution of the Stokes equations. Then, unlike the textbook problem of a conduit of fixed shape, in \S\ref{sec:plate} we couple the equations from \S\ref{sec:lubri} to the governing equations of Kirchhoff--Love plate theory. On the basis of the leading-order solutions for the streamwise velocity component and for the plate deformation, in \S\ref{sec:dp-q} a fitting-parameter-free expression for the flow rate--pressure drop relation for a long, shallow microchannel is derived. In \S\ref{sec:expt}, the analytical expressions derived are illustrated for a range of values of the dimensionless parameters, and shown to compare favorably with experimental measurements. Finally, conclusions are stated in \S\ref{sec:concl}. In Appendix~\ref{sec:stiff_topwall}, some explicit formul\ae\ for the velocity profile and the pressure as functions of the axial coordinate are presented for the special case of a microchannel with a ``stiff'' top wall. Meanwhile, Appendix~\ref{app:notshallow} includes further mathematical details regarding the effect of the lateral sidewalls on our analysis.


\section{Lubrication approximation for shallow deformable channels}
\label{sec:lubri}
\begin{figure}
	\centering
	\vspace{5mm}
	\includegraphics[width=0.75\textwidth]{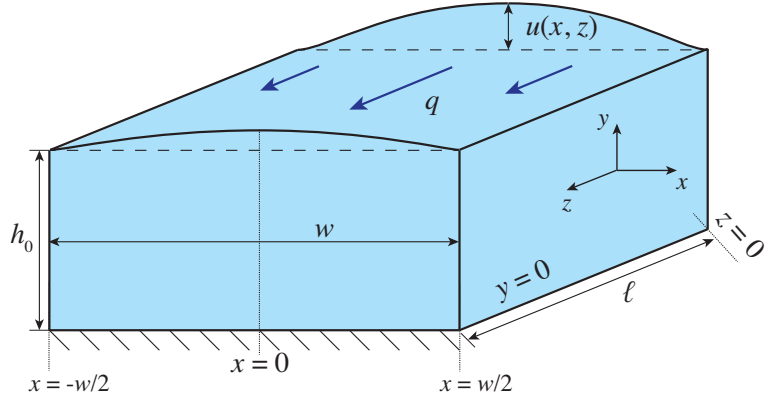}
	\caption{Three-dimensional schematic of the deformed geometry of a channel of length $\ell$ with an initially rectangular cross-section ($w\times h_0$). The volumetric flow rate is denoted by $q$, and $u(x,z)$ is the top wall's deformation (the remaining walls are assumed rigid). Although not shown in this schematic, the top wall has a thickness $t$.}
	\label{fig:schematic}
\end{figure}

We consider a channel of length $\ell$, width $w$ and height $h$, where $h \ll w \ll \ell$. The upper wall of the channel is soft and deformable, as is the case when a rigid channel is sealed by a thin elastic film. The flow is in the positive $z$-direction. Due to the normal stresses from the flow on the walls, the (soft) top wall of the channel deforms out of the $(x,z)$-plane in the positive $y$-direction, so that the steady shape of the channel's top wall is given by $y = h(x,z) = h_0 + u(x,z)$, as shown in figure~\ref{fig:schematic}(\textit{a}). For the moment, we make no assumption on the magnitude of the displacement, however, we expect that $u>0$ and $u\ll w$ for the types of problems of interest herein.

Now, consider the incompressible flow of a Newtonian fluid of density $\varrho$ and viscosity $\mu$. As is typical for microfluidic devices \citep{SSA04,SQ05,B08}, we are interested in the limit in which the effective Reynolds number is small, \textit{i.e.}, $\epsilon Re = (h_0/\ell)\varrho\mathcal{V}_c h_0/\mu\ll1$, where $\mathcal{V}_c$ is a characteristic axial velocity and $\epsilon = h_0/\ell \ll 1$. In this limit, fluid inertia is negligible compared to viscous stresses, and the flow is governed by the Stokes equations \citep{HB83,B08}:%
\refstepcounter{equation}
$$
\bm{0} = - \bnabla p + \mu \bnabla^2\bm{v},\qquad \bnabla\bcdot\bm{v} = 0,
  \eqno{(\theequation{\mathit{a},\mathit{b}})}
\label{eq:stokes}
$$
where $\bm{v} = (v_x,v_y,v_z)$ and body forces have been neglected. The flow is also subject to the no-slip and no-penetration boundary conditions along the walls of the channel:
\begin{equation}
\bm{v} = \bm{0}\quad\text{at}\quad \begin{cases} y = 0,\, h(x,z),\\ x = \pm w/2. \end{cases}
\label{eq:all_bc}
\end{equation}
Note that the velocity vanishes along $y=h$, rather than being $v_x = dh/dt$, etc., because we have already assumed the fluid flow and the structural deformation to be steady ({\it i.e.}, independent of time).

Let us introduce the dimensionless variables
\begin{multline}
x = wX,\quad y = h_0Y,\quad z = \ell Z,\quad h(x,z) = h_0 H(X,Z),\quad u(x,z) = u_\mathrm{c} U(X,Z),\\ {v}_x(x,y,z) = \frac{\epsilon}{\delta} {\mathcal{V}}_{\mathrm{c}} {V}_X(X,Y,Z),\quad {v}_y(x,y,z) = \epsilon {\mathcal{V}}_{\mathrm{c}} {V}_Y(X,Y,Z),\\ {v}_z(x,y,z) = {\mathcal{V}}_{\mathrm{c}} {V}_Z(X,Y,Z),\quad p(x,y,z) = \Delta p\, P(X,Y,Z),
\label{eq:dimless_vars}
\end{multline}
where $\Delta p$, $u_\mathrm{c}$ and $\mathcal{V}_\mathrm{c}$ are a characteristic pressure scale, a characteristic top wall deformation scale, and a characteristic axial velocity scale, respectively; we have also defined $\delta := h_0/w$ and $\epsilon := h_0/\ell$. The characteristic top wall deformation $u_\mathrm{c}$ is not an independent parameter and will be determined below, upon scaling the governing equations. First, consider the continuity equation (\ref{eq:stokes}\textit{b}), which becomes
\begin{equation}
\frac{\partial V_X}{\partial X} + \frac{\partial V_Y}{\partial Y} + \frac{\partial V_Z}{\partial Z} = 0.
\label{eq:divV_dimless}
\end{equation}
By convention, we sought a balance in the continuity equation, which is how the velocity scales in \eqref{eq:dimless_vars} were set.

A microchannel can be operated in one of two regimes: either the pressure drop $\Delta p = p(0) - p(\ell)$ is imposed, where we assume that the outlet is always open to the atmosphere so we set $p(\ell) = 0$ (then, the inlet pressure is $p(0) = \Delta p$), or the flow rate $q$ is imposed, which is equivalent to imposing the mean flow velocity $\langle v_z \rangle \propto q/(h_0w)$. In both cases, $\Delta p$ and $\mathcal{V}_\mathrm{c}$ are not independent scales but coupled by the flow physics. In the first case, the characteristic velocity scale is set by pressure drop $\Delta p$: $\mathcal{V}_\mathrm{c} = h_0^2 \Delta p/(\mu\ell)$; in the second case, the pressure drop is set by the characteristic axial velocity scale $\mathcal{V}_\mathrm{c} = \langle v_z \rangle$: $\Delta p =  \mu \langle v_z \rangle\ell/h_0^2$. In either case, (\ref{eq:stokes}\textit{a}) adopts the same form:
\begin{subequations}\label{eq:stokes_dimless3}\begin{align}
\epsilon^2\delta^2 \frac{\partial^2 V_X}{\partial X^2} +  \epsilon^2\frac{\partial^2 V_X}{\partial Y^2} + \epsilon^4 \frac{\partial^2 V_X}{\partial Z^2} &= \delta^2 \frac{\partial P}{\partial X},\label{eq:stokes_dimless3_X}\\
\epsilon^2\delta^2 \frac{\partial^2 V_Y}{\partial X^2} + \epsilon^2\frac{\partial^2 V_Y}{\partial Y^2} + \epsilon^4 \frac{\partial^2 V_Y}{\partial Z^2} &= \frac{\partial P}{\partial Y},\\
\delta^2 \frac{\partial^2 V_Z}{\partial X^2} + \frac{\partial^2 V_Z}{\partial Y^2} + \epsilon^2 \frac{\partial^2 V_Z}{\partial Z^2} &= \frac{\partial P}{\partial Z}.\label{eq:stokes_dimless3_Z}
\end{align}\end{subequations}

Above, we assumed that the channel is long and thin, which leads to a natural ordering of the small (dimensionless) parameters:
\begin{equation}\label{eq:ordering}
0 < \epsilon \ll \delta \ll 1.
\end{equation}
We are interested in the leading-order asymptotic behavior under the ordering \eqref{eq:ordering}. From \eqref{eq:stokes_dimless3}, we have%
\refstepcounter{equation}
$$
\frac{\partial^2 V_{Z}}{\partial Y^2} = \frac{\partial P}{\partial Z},\qquad
0 = \frac{\partial P}{\partial Y},\qquad  0 = \frac{\partial P}{\partial X},
  \eqno{(\theequation{\mathit{a},\mathit{b},\mathit{c}})}
\label{eq:leading_order}
$$
which is the familiar lubrication approximation.
From (\ref{eq:leading_order}\textit{b},\textit{c}), we first infer that the leading-order pressure does not depend on the cross-sectional coordinates $(X,Y)$. Then, after integrating (\ref{eq:leading_order}\textit{a}) and enforcing the no-slip boundary condition from \eqref{eq:all_bc}, \textit{i.e.}, $V_{Z}(X, 0, Z) = V_{Z}\big(X, H(X,Z), Z\big) = 0$ $\forall X,Z$, the ``standard'' lubrication theory result for the axial velocity follows:
\begin{equation}
V_{Z}(X,Y,Z) = -\frac{1}{2}\frac{\rd P}{\rd Z} \big( H(X,Z)-Y \big)Y.
\label{eq:VZ00}
\end{equation}
We remind the reader that $\rd P/\rd Z < 0$ in our convention. This leading-order axial velocity does not satisfy the no-slip boundary condition at the side walls ($X=\pm 1/2$). Appendix~\ref{app:notshallow} discusses the effect of the lateral side walls, showing their effect is indeed small for $\delta\ll1$.

From the velocity scalings given in \eqref{eq:dimless_vars}, it is clear that the cross-sectional velocity components are much smaller than the axial one. Thus, to the leading order in the two small parameters, the velocity field in this lubrication approximation is dominated by the axial component $V_Z$. 
Terms beyond the leading-order ones can be computed in the standard way, by positing a regular perturbation expansion in the \emph{two} small parameters $\epsilon$ and $\delta$.

Next, we wish to relate the volumetric flow rate though the channel, 
\begin{equation}
Q := \int_{-1/{2}}^{+1/{2}} \int_{0}^{H(X,Z)} V_Z(X,Y,Z)\, \rd Y \rd X,
\label{eq:flow_rate}
\end {equation}
to the streamwise gradient of the pressure $\rd P/\rd Z$. Since the flow is steady, $Q$ is constant. Then, substituting $V_Z$ from \eqref{eq:VZ00} into \eqref{eq:flow_rate} and integrating over $Y$, we get (to the leading order in $\delta$ and $\epsilon$) the ``standard'' lubrication theory relation
\begin{equation}
Q = -\frac{1}{12} \frac{\rd P}{\rd Z}\int_{-1/{2}}^{+1/{2}} H(X,Z)^3 \,\rd X.
\label{eq:flow_rate2}
\end{equation}
Thus, for a given shape of the top wall of the channel, we can find $Q$ from \eqref{eq:flow_rate2}. Within the lubrication approximation, the first correction in $\delta$ to \eqref{eq:flow_rate2} can be computed as shown in Appendix~\ref{app:notshallow}.

\section{Shape of the deformed channel}
\label{sec:plate} 

In terms of the dimensionless variables from \eqref{eq:dimless_vars}, the channel shape can thus be expressed as
\begin{equation}
H(X,Z) = 1 + \beta U(X,Z),
\label{eq:topwall_dimless}
\end{equation}
where we have set $\beta := u_\mathrm{c}/h_0$. Here, the dimensionless group $\beta$ controls the compliance of the top wall: for $\beta\ll1$, the top wall is stiff (equivalently, its deformation is small compared to the undeformed height), and for $\beta \gg 1$ it is soft (equivalently, its deformation is large compared to the undeformed height). Within the lubrication approximation, we do not need to make any assumptions on the smallness of $\beta$ at this stage, however, we expect $\beta>0$ and $\beta\delta\ll1$ as stated above in \S\ref{sec:lubri} (see also Appendix~\ref{app:notshallow}).

If the maximum displacement $\max_{x,z}|u(x,z)|$ of the top wall can be assumed small compared to its thickness $t$, and its thickness $t$ is small compared to its width $w$, then, from the theory of linear elasticity, we know that the steady-state displacement $u(x,z)$ satisfies the Kirchhoff--Love equation for isotropic quasi-static bending of a plate under a transverse load due to the fluid pressure \citep{L88,TWK59,LL86}:
\begin{equation}
B\bnabla_{\|}^2\bnabla_{\|}^2 u = p,
\label{eq:KL_plate}
\end{equation}
where $B = Et^3/[12(1-\nu^2)]$ is the bending energy (flexural rigidity) of the plate, $E$ is the material's Young's modulus, and $\nu$ is the Poisson ratio of the material; $\bnabla_{\|}^2$ is the Laplacian operator in the $(x,z)$ coordinates (tangent to the base  flow). Equation \eqref{eq:KL_plate} models only bending of the plate, with stretching being assumed negligible; if stretching is judged to be significant, then the F\"oppl--von K\'{a}rm\'{a}n equations can be employed \citep{TWK59,LL86}. 

Using the dimensionless variables from \eqref{eq:dimless_vars}, equation \eqref{eq:KL_plate} becomes
\begin{equation}
\delta^4\frac{\partial^4 U}{\partial X^4} + 2\epsilon^2\delta^2\frac{\partial^4 U}{\partial X^2\partial Z^2} + \epsilon^4\frac{\partial^4 U}{\partial Z^4} = \frac{h_0^4 \Delta p}{B u_\mathrm{c}} P.\label{eq:beam}
\end{equation}
To obtain a dominant balance in \eqref{eq:beam}, we must take $h_0^4\Delta p/(B u_\mathrm{c}) = \delta^{4}$ or, specifically, this balance sets the characteristic deformation scale to be
\begin{equation}
u_\mathrm{c} = \frac{h_0^4\Delta p}{B \delta^4} = \frac{w^4\Delta p}{B}.
\label{eq:uc}
\end{equation}
Hence, $\beta \equiv u_\mathrm{c}/h_0 = h_0^3\Delta p/(B \delta^4)$; typical values of $\beta$ are estimated in table~\ref{tb:params1} below. With $u_\mathrm{c}$ determined, equation \eqref{eq:beam} gives, to the leading order,
\begin{equation}
\frac{\partial^4 U}{\partial X^4} = P(Z).
\label{eq:beam2}
\end{equation}
Physically, this asymptotic limit can be interpreted as follows: Since $w\ll \ell$, the variation of the height of the channel in the streamwise direction occurs over a much longer length scale than the variation of the height in any cross-section. In other words, the deflection of any infinitesimal slice (perpendicular to the flow-wise axial direction) of the top wall does not affect infinitesimal slices nearby. Thus, the Kirchhoff--Love plate theory reduces to the Euler--Bernoulli beam theory for each infinitesimal spanwise slice of the top wall. This limit should be contrasted to the case when $\delta \sim \epsilon$ (and the elastic top wall is \emph{not} clamped along $X=\pm1/2$) in which case bending (and, potentially, tension) in the \emph{flow-wise} direction dominate \citep{GMV17}.

Of course,  at this order in the asymptotic analysis, the beam equation \eqref{eq:beam2} cannot satisfy the boundary conditions along $Z=0$ or $Z=1$ (the channel's entrance and exit planes). In experiments, the plate is also clamped along $Z=0$ and $Z=1$. From \eqref{eq:beam}, it is evident that a boundary layer calculation can be done by the rescaling $Z\mapsto (\epsilon/\delta)Z$, which keeps all terms on the left-hand side of \eqref{eq:beam}. Thus, any corrections due to clamping at $Z=0$ and $Z=1$ are localized to layers of width $\mathcal{O}(\epsilon/\delta)$, which are a small correction to the system's response in the central part of the channel, given our assumed ordering of small parameters \eqref{eq:ordering}.

We take the plate to be clamped at $X=\pm{1}/{2}$, so \eqref{eq:beam2} is subject to the boundary conditions
\begin{equation}\label{eq:beam_bc}
U\left(\pm{1}/{2}, Z\right) = 0,\qquad \left. \frac{\partial U}{\partial X} \right|_{X=\pm1/{2}} = 0.
\end{equation}
Integrating equation \eqref{eq:beam2} four times with respect to $X$ and enforcing the boundary conditions from \eqref{eq:beam_bc}, we find that 
\begin{equation}
U(X,Z) = \frac{1}{24} P(Z) \left( X + \tfrac{1}{2} \right)^2 \left( X - \tfrac{1}{2} \right)^2.
\label{eq:U_of_XZ}
\end{equation}
Note that, because $u_c$ is set by bending via \eqref{eq:uc}, we cannot take a limit as a parameter in \eqref{eq:U_of_XZ} vanishes and recover the case of a rigid channel (\textit{i.e.}, $U=0$). 

The top-wall displacement given by \eqref{eq:U_of_XZ} is a \emph{quartic} polynomial of the cross-sectional coordinate $X$, which corresponds to our assumption that the system is in a bending-dominated regime. Nevertheless, it should also be noted that a (different) quartic profile is consistent with recent experimental measurements of a similar model system \citep{DHTJ17}, in which bending and pre-tension can \emph{both} be considered significant. In other literature \citep{GEGJ06,CTS12,OYE13,RS16}, however, a {parabolic} (\emph{quad\-ratic}) profile has often been assumed, even without making a distinction between (or an evaluation of) bending-dominated versus stretching-dominated deformation.

It is often of interest to relate the cross-sectionally averaged displacement to the maximum displacement, since the latter is easier to measure experimentally. From \eqref{eq:U_of_XZ} we find that
\refstepcounter{equation}
$$
  \overline{U}(Z) := \int_{-1/2}^{+1/2} U(X,Z) \,\rd X = 
  \frac{8}{15} U_\mathrm{max}(Z),\qquad U_\mathrm{max}(Z) := \frac{1}{384}P(Z).
   \eqno{(\theequation{\mathit{a},\mathit{b}})}
  \label{eq:ubar_max}
$$
In passing, we note that the prefactor in (\ref{eq:ubar_max}\textit{a}), when computed on the basis of the parabolic displacement approximation is $2/3$ rather than $8/15$; a 25\% difference. The predicted linear scaling of the maximum and cross-sectionally averaged displacements with the pressure is also consistent with estimates of the static deflection of plates by \cite{TWK59}. 

\section{The flow rate--pressure drop relation}
\label{sec:dp-q}

To find the flow rate through the deformed channel, we must evaluate the integral in \eqref{eq:flow_rate2} using the top-wall shape stipulated in \eqref{eq:topwall_dimless}:
\begin{equation}
Q = -\frac{1}{12} \frac{\rd P}{\rd Z}\int_{-1/{2}}^{+1/{2}} [1+\beta U(X,Z)]^3 \,\rd X.\\
\label{eq:flow_rate_int_h}
\end{equation}
This integral can be evaluated upon substituting the expression for $U$ from \eqref{eq:U_of_XZ} to find
\begin{equation}
Q = -\frac{1}{12}\frac{\rd P}{\rd Z}\left[1 + \frac{\tilde\beta}{10} P(Z) + \frac{\tilde\beta^2}{210} P(Z)^2 + \frac{\tilde\beta^3}{12,012} P(Z)^3 \right], 
\label{eq:q_of_dpdz}
\end{equation}
where $\tilde\beta := \beta/24 = h_0^3\Delta p/(24 B \delta^4)$. In passing, we note that \eqref{eq:q_of_dpdz} is a relationship of the form $Q = -\sigma(P)\rd P/\rd Z$, for some function $\sigma$ that comes about from solving the elasticity problem, as discussed by \citet{RK72}.

For a constant imposed flow rate $Q$, \eqref{eq:q_of_dpdz} is a separable first-order ordinary differential equation for the pressure distribution $P(Z)$ subject to $P(1) = 0$, whence
\begin{equation}
Q = \frac{P(Z)}{12(1-Z)}\left[ 1 + \frac{\tilde\beta}{20}P(Z) + \frac{\tilde\beta^2}{630} P(Z)^2 + \frac{\tilde\beta^3}{48,048} P(Z)^3 \right].
\label{eq:q_of_p}
\end{equation}
As $Q$ is constant, \eqref{eq:q_of_p} is a polynomial, whose real positive root gives the dimensionless fluid pressure $P$ as a function of the streamwise coordinate $Z$ and the various dimensionless parameters.\footnote{Though it is possible to obtain the roots of the quartic polynomial in \eqref{eq:q_of_p} by Ferrari's procedure \cite[see, \textit{e.g.},][\S3.8.3]{AS72}, the explicit formula for the sought positive real root is far too long to be presented here. It can, however, be  found easily using {\sc Mathematica}.}
For $\tilde{\beta} = 0$ (an ``infinitely stiff'' plate), there is no deformation, and \eqref{eq:q_of_p} reduces to the classical lubrication-theory result for a rectangular channel.

Since the last two terms in the bracket in \eqref{eq:q_of_p} have fairly large denominators, they can be neglected for small as well as moderate values of $\tilde\beta P(Z)$, the consequences of which are discussed in Appendix~\ref{sec:stiff_topwall}. Obviously, for larger values of $\tilde\beta P(Z)$ these higher-order terms become important.

\subsection{Flow-rate-controlled versus pressure-drop-controlled problems}

For a pressure-driven flow, $\Delta P=1$ under the nondimensionalization $p = \Delta p\, P$; hence $P(0) = 1$. Then, the flow rate--pressure drop relation \eqref{eq:q_of_p} can be evaluated at $Z=0$ to obtain
\begin{equation}
12 Q = 1 + \frac{\tilde\beta}{20} + \frac{\tilde\beta^2}{630} + \frac{\tilde\beta^3}{48,048},
\label{eq:q_of_dp}
\end{equation}
which is a {universal} dimensionless relationship valid for {any} applied $\Delta p$. 

For a flow-rate controlled situation, $Q=1$ under the nondimensionalization $p = \mu \ell q/(h_0^3w) P$; hence $P(0) = \Delta P$. Then, the flow rate--pressure drop relation \eqref{eq:q_of_p} can be evaluated at $Z=0$ to obtain
\begin{equation}
12 = \Delta P\left[1 + \frac{\tilde\beta}{20}\Delta P + \frac{\tilde\beta^2}{630}(\Delta P)^2 + \frac{\tilde\beta^3}{48,048} (\Delta P)^3\right],
\label{eq:q_of_dp2}
\end{equation}
which is a {universal} dimensionless relationship valid for {any} imposed $q$.

Equations \eqref{eq:q_of_dp} and \eqref{eq:q_of_dp2} can also be interpreted as consistency checks. For a given experiment (or numerical simulation) with an imposed (dimensionless) flow rate $Q$ or an imposed dimensionless pressure drop $\Delta P$, the value of $\tilde{\beta}$ corresponding to the system must satisfy either the \eqref{eq:q_of_dp} or  \eqref{eq:q_of_dp2}.

\subsection{Comparison with previous models}
\label{sec:previous_models}

Equation~\eqref{eq:q_of_p}, which contains no fitting parameters, can be compared to equation (10) of \citet{GEGJ06}, which incorporates a fitting parameter $\alpha$ and has the dimensional form:
\begin{equation}
q = \frac{h_0^3 w\, p(z)}{12\mu(\ell-z)} \left[ 1 + \frac{3}{2}\frac{\alpha w}{Eh_0} p(z) + \left(\frac{\alpha w}{Eh_0}\right)^2 p(z)^2 + \frac{1}{4}\left(\frac{\alpha w}{Eh_0}\right)^3 p(z)^3\right].
\label{eq:q_p_gervais_dim}
\end{equation}
Equation \eqref{eq:q_p_gervais_dim} was derived by \citet{GEGJ06} under the assumption that the top wall is thick, and the strains within it, $u/w$, are proportional to $p/E$. The proportionality constant is $\alpha$, and it is unknown {\it a priori}. Although \citet{GEGJ06} only discuss the case of a thick top wall that behaves like an elastic half-space, \eqref{eq:q_p_gervais_dim} has been used by others \citep[see, {\it e.g.},][]{HUZK09,CTS12,RDC17} for microchannels with much thinner top walls that behave as plates. Consequently, when \eqref{eq:q_p_gervais_dim} is fit to flow pressure--drop data from microchannels with plate-like top walls, a dependence $\alpha \propto (t/w)^3$ is observed \citep{RDC17}.

For the bending-dominated analysis presented herein, we found [similarly to \citet{GEGJ06}] that $u/w\propto p/E$, but on the basis of plate theory. Consequently, it is possible to compare our flow rate--pressure drop relation \eqref{eq:q_of_p} to a generic relation such as \eqref{eq:q_p_gervais_dim} since they are both based on the top wall displacement scaling with the fluid pressure, {\it i.e.}, $u/w\propto p/E$. In order to discuss the similarities and differences between these two relations, we note that the dimensional form of \eqref{eq:q_of_p} is
\begin{multline}
q = \frac{h_0^3w\,p(z)}{12\mu(\ell-z)}\left[ 1 + \frac{1}{480}\frac{w^4}{B h_0}p(z) + \frac{1}{362,880}\left(\frac{w^4}{B h_0}\right)^2 p(z)^2\right. \\ \left. + \frac{1}{664,215,552}\left(\frac{w^4}{B h_0}\right)^3 p(z)^3 \right].
\label{eq:q_of_p_dim}
\end{multline}
Clearly, the quantity in the brackets on the right-hand side of \eqref{eq:q_p_gervais_dim} can be 
compared with the quantity in the brackets on the right-hand side of \eqref{eq:q_of_p_dim}. The caveat here is that we assume that \eqref{eq:q_p_gervais_dim} can be applied to the bending-dominated top wall deformation regime. In this comparison, then, the role of the parameter $\alpha$ is to absorb the necessary details of the displacement profile and its scaling with various material and geometric constants. For example, the second terms in the brackets of \eqref{eq:q_p_gervais_dim} and \eqref{eq:q_of_p_dim} can be made to agree if we choose $\alpha = \tfrac{1}{60} (w/t)^3 (1-\nu^2)$. We cannot make all terms agree because \eqref{eq:q_of_p_dim} is based on the cross-sectional displacement profile $u(x,z)$ derived in \S\ref{sec:plate},  while \citet{GEGJ06} postulated that $u/w = \alpha p/E$. 

\citet{RS16} and \citet{RDC17} also observed that $\alpha$ need not be a fitting parameter but, instead, must be related to the various geometric and elasticity constants involved in the problem. A  difference between the approaches of \citet{RS16} and \citet{RDC17} and ours is that their derivations are based on correlating the maximum of the cross-sectionally-averaged displacement to the pressure via various approximations/assumptions. In particular, \citet{RS16} make the assumption of a parabolic top wall deflection, which does not appear to be well-substantiated by their measurements. Consequently, the (dimensional) ``compliance parameters,'' $f_\mathrm{p}$ introduced by \citet{RS16} and $\Omega$ introduced by \citet{RDC17}, are derived differently from (the dimensionless) $\tilde\beta$ introduced above on the basis of the perturbation expansion of the coupled fluid--structure interaction problem. 

More importantly, however, \citet{RS16} based their analysis on the shell-stretching correlation $u/w \propto (p/E)^{1/3}$, which yields a model that cannot be compared to those based on linearly proportional displacement and pressure (such as those discussed above). On the other hand, the model flow rate--pressure drop relation of \citet{RDC17} was based on a plate-bending correlation, so it can be directly compared to \eqref{eq:q_of_p_dim}. Unfortunately, there are misprints in the derivation of equation~(12) of \citet{RDC17}, {\it i.e.}, the model flow rate--pressure drop relation therein. Specifically, the compliance parameter $\Omega$ as given in equation~(10) of \citet{RDC17} should read $\Omega = \frac{1}{99}E^{-1}(w/h_0)(w/t)^3(1-\nu^2)$ and the flow rate--pressure drop relation from equation~(12) of \citet{RDC17} should read
\begin{equation}
q = \frac{h_0^3 w\, p(z)}{12\mu(\ell-z)} \left[ 1 + \frac{3}{2} \Omega p(z) + \Omega^2 p(z)^2 + \frac{1}{4}\Omega^3 p(z)^3\right].
\label{eq:q_p_raj2_dim}
\end{equation}
On comparing \eqref{eq:q_p_raj2_dim} to \eqref{eq:q_p_gervais_dim}, it follows that $\alpha = \frac{1}{99}(w/t)^3(1-\nu^2)$. Unsurprisingly, this comparison yields a constant $\alpha$, which is no longer a fitting parameter. The latter prediction differs from ours above in the prefactor $\frac{1}{99} \approx 0.011$ versus $\frac{1}{60} \approx 0.017$. Furthermore, as was the case with \eqref{eq:q_p_gervais_dim}, \eqref{eq:q_p_raj2_dim} cannot be made to agree with \eqref{eq:q_of_p_dim} because \eqref{eq:q_p_raj2_dim} was derived following the approach of \cite{GEGJ06}, while \eqref{eq:q_of_p_dim} was derived using the actual cross-sectional displacement profile $u(x,z)$.

\section{Illustrated examples and comparison with experiments}
\label{sec:expt}

\begin{figure}
	\centering
	\vspace{5mm}
	\includegraphics[width=\textwidth]{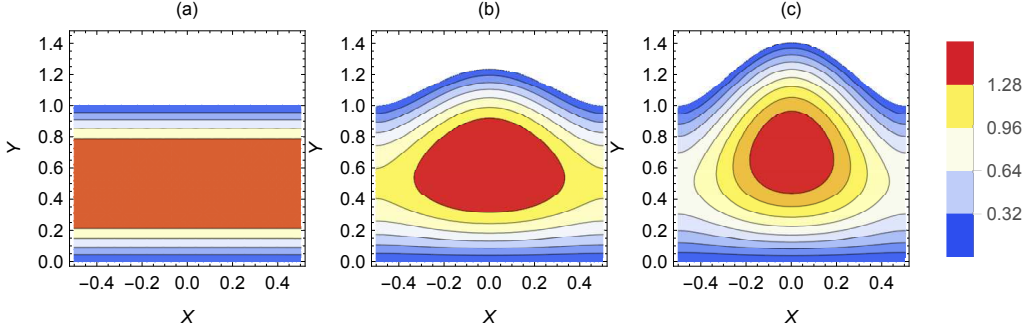}
	\caption{Contours of the axial velocity, $V_Z$, as a function of the $(X,Y)$ coordinates in the $Z=0.25$ cross-section of the channel with $Q=1$ and for different $\tilde{\beta}$ values, computed from \eqref{eq:VZ00} with the pressure given by the appropriate real root of \eqref{eq:q_of_p} and the channel height given by \eqref{eq:topwall_dimless} and \eqref{eq:U_of_XZ}. (\textit{a}) Solid top wall ($\tilde\beta =0$). (\textit{b}) Elastic top wall with $\tilde\beta = 0.5$. (\textit{c}) Elastic top wall with $\tilde\beta = 1$.} 
	\label{fig:v_compare}
\end{figure}

First, consider the leading-order axial velocity component \eqref{eq:VZ00}. Clearly, because $P=P(Z)$ and $H=H(X,Z)$, the deformation of the channel's top wall introduces dependence upon \emph{both} of the coordinates (\textit{i.e.}, $X$ and $Z$) that are not present in the leading-order velocity in a rigid channel, namely $V_Z(X,Y,Z)=6Q(1-Y)Y$. The effect of the top wall's elasticity on the axial velocity is illustrated in figure~\ref{fig:v_compare}. Specifically, it is evident that the cross-sectional shape's deformation introduces variability in the $X$-direction, which leads to faster flow near the channel's centerline (compared to the rest of the cross-section), consistent with the numerical simulations shown in figure~3 of \citep{GEGJ06}.

\begin{figure}
	\centering
	\vspace{5mm}
	\includegraphics[width=\textwidth]{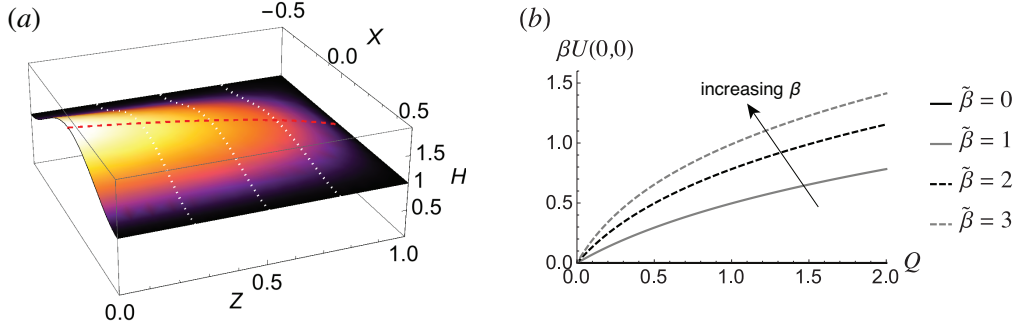}
	\caption{(\textit{a}) $H(X,Z)$ computed from \eqref{eq:U_of_XZ} with $P(Z)$ computed from  \eqref{eq:q_of_p}, for $\tilde{\beta} = 2$ and $Q=1$; thick dashed line represents $1+\beta U_\mathrm{max}(Z)$, thin dotted lines are a guide to the eye. (\textit{b}) Maximum displacement, \textit{i.e.}, $H(0,0) - 1 = \beta U(0,0)$ calculated from \eqref{eq:U_of_XZ} with $P(Z)$ computed from \eqref{eq:q_of_p}, as function of the flow rate $Q$ and different values of $\tilde{\beta}$.} 
	\label{fig:displacement}
\end{figure}

Second, the channel's top wall deformation is illustrated in figure~\ref{fig:displacement}. In dimensionless variables, figure~\ref{fig:displacement}(\textit{a}) shows a surface plot of the top-wall displacement as a function of the spatial coordinates, while figure~\ref{fig:displacement}(\textit{b}) shows the maximum displacement of the top wall as a function of the flow rate $Q$. The shapes of the latter curves agree qualitatively with those in figures~5 and 7 of \citet{GEGJ06}, respectively.

\begin{figure}
	\centering
	\vspace{5mm}	
	\includegraphics[height=0.26\textwidth]{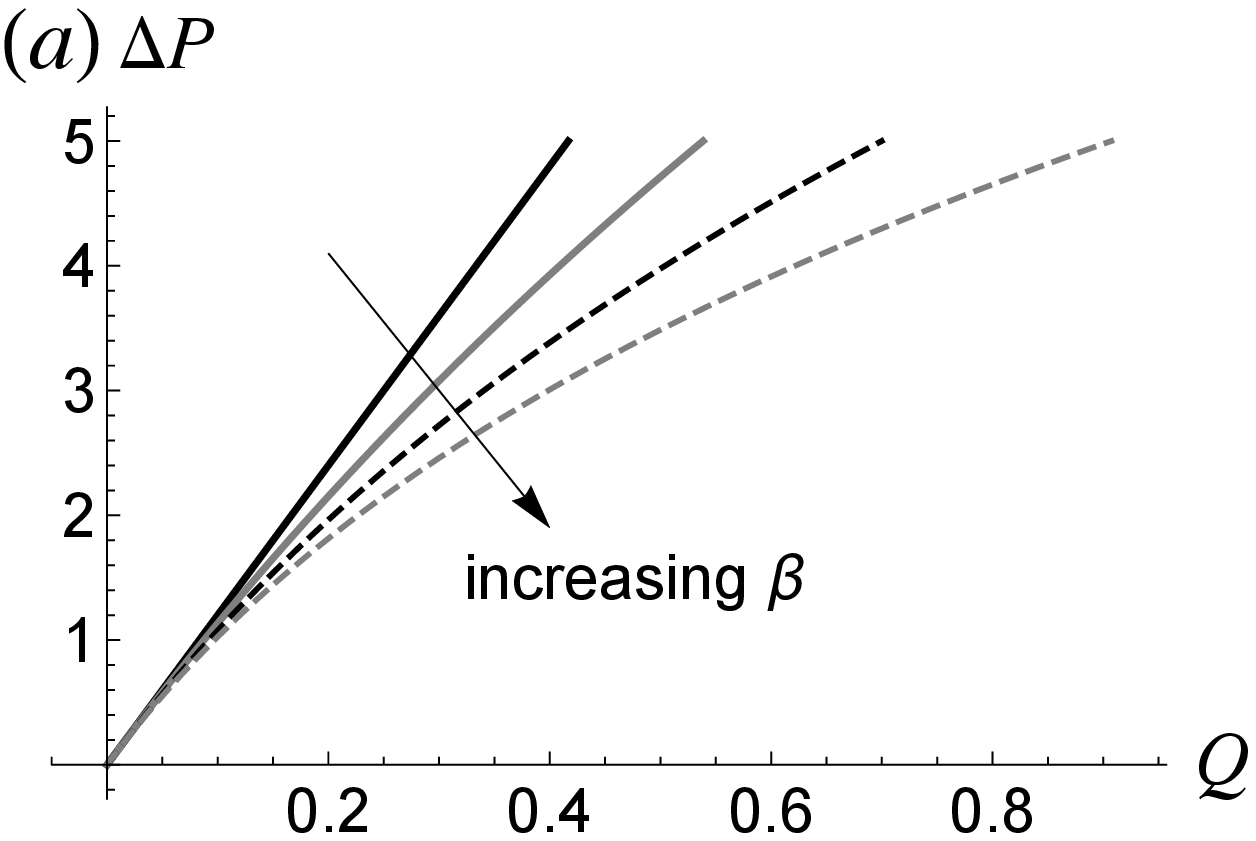}\hfill
	\includegraphics[height=0.26\textwidth]{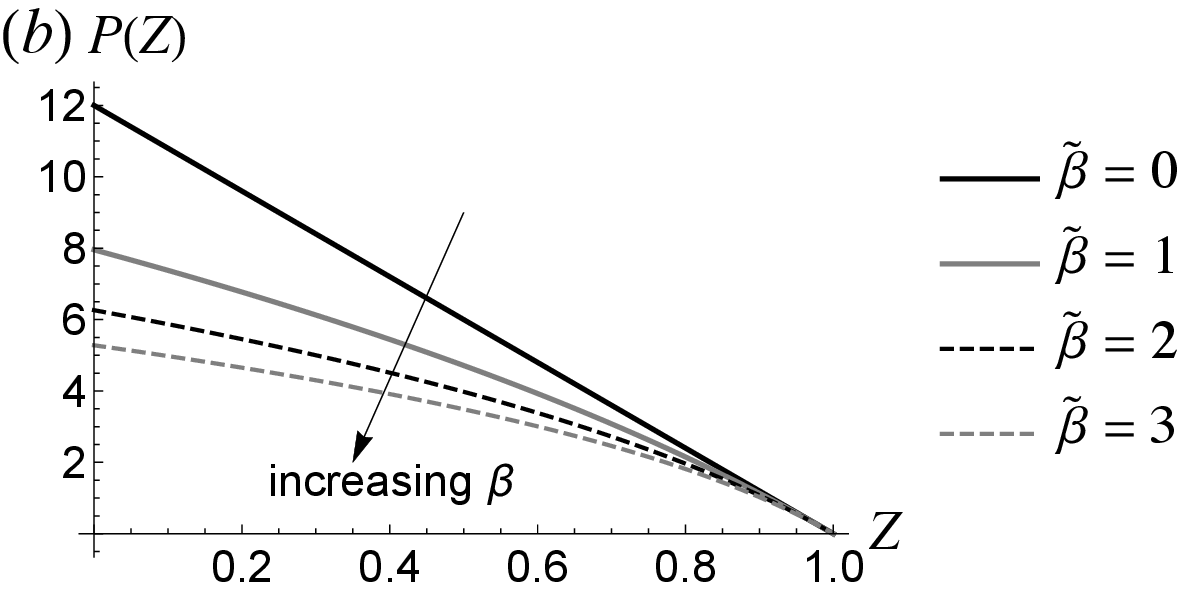}
	\caption{(\textit{a}) Pressure drop across the channel, $\Delta P = P(0)$, computed via \eqref{eq:q_of_p}, as a function of the flow rate $Q$ for different values of $\beta$.  (\textit{b}) Pressure $P$ as a function of the axial coordinate $Z$ computed by inverting \eqref{eq:q_of_p}; $Q=1$.} 
	\label{fig:pressure-flow}
\end{figure}

Third, the \emph{nonlinear} flow rate--pressure drop curves for a deformable channel are illustrated in figure~\ref{fig:pressure-flow}(\textit{a}) for different values of $\tilde\beta$. Their shapes agree qualitatively with \citet{GEGJ06} (figure~9 therein). From figure~\ref{fig:pressure-flow}(\textit{a}), it is clear that, for $\tilde\beta \ne 0$ (a deformable channel), the value of $Q$ for a given $\Delta P$ can be significantly larger as $\beta$ is increased, {\it i.e.}, for softer channels. Meanwhile, figure~\ref{fig:pressure-flow}(\textit{b}) shows the pressure in the microchannel as a function of the axial coordinate using the implicit expression \eqref{eq:q_of_p}. Clearly, a contribution of flow--elasticity coupling is to make the pressure distribution in the channel {nonlinear}. Furthermore, since the channel cross-sectional area increases due to the top wall's deformation, a smaller pressure gradient is needed to achieve the same volumetric flow rate.

Next, we present comparisons between the predictions of our theoretical analysis and experimental results previously published in the literature. To this end, table~\ref{tb:params1} summarizes the values of the physical parameters in the flow-rate-controlled PDMS microchannels experiment, which best matches the assumptions of our theoretical developments, labeled ``S4'' in \citep{OYE13}. The ranges of $\beta$ and $\tilde\beta$, which were computed using \eqref{eq:q_of_dp}, correspond to the range of flow rates considered by \citet{OYE13}, namely $q=10$ to $50$ mL/min, or $Q=0.1$ to $0.15$. From table~\ref{tb:params1}, which also shows the values of the dimensionless quantities that we have introduced, we see that $\delta \approx 0.144$ while $\epsilon \approx 0.02$, which satisfies our assumed asymptotic ordering: $0<\epsilon\ll\delta\ll1$. Regarding our assumption that plate-bending theory captures the mechanics of the top-wall deformation, we note, from table~\ref{tb:params1}, that $t/w = 0.11 \ll 1$, while $\max_{x,z} |u(x,z)|/t \le 0.43 < 1$ based on the results of \citet{OYE13} (with $0.43$ being the value at the largest flow rate considered therein).

\begin{table}
\begin{center}
\def~{\hphantom{0}}
  \begin{tabular}{l@{\hskip 8pt}l@{\hskip 8pt}l@{\hskip 8pt}l@{\hskip 8pt}l@{\hskip 8pt}l@{\hskip 8pt}l@{\hskip 8pt}l@{\hskip 8pt}l@{\hskip 8pt}l@{\hskip 8pt}l@{\hskip 8pt}l}
    $h_0$ & $w$ & $\ell$ & $\Delta p$ & $t$ & $E$ & $\nu$ & $B$ & $\delta$ & $\epsilon$ & $\beta$ & $\tilde\beta$\\  
    (mm) & (mm) & (mm) & (kPa) & (mm) & (MPa) & & ($\mu$J) & & & & \\[3pt]    
    $0.244$ & 1.7 & 15.5 & 1--5 & $0.2$ & $\approx1.6$ & 0.499 & $\approx1.60$ & 0.144 & 0.0244 & 24.5--97 & 1--4
  \end{tabular}  
  \caption{Values of the physical parameters for the flow-rate-controlled PDMS microchannels experimental system S4 of \citet{OYE13}, which most closely matches the assumptions of our theoretical developments.}
\label{tb:params1}
\end{center}
\end{table}

\begin{figure}
	\centering
	\vspace{5mm}	
	\includegraphics[height=0.3\textwidth]{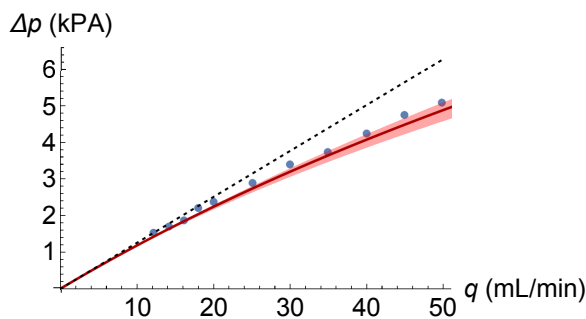}
	\caption{Dimensional flow rate--pressure drop relationship: comparison between our asymptotic theory (solid curve) [{\it i.e.}, \eqref{eq:q_of_p_dim} evaluated at $z=0$], including a shaded trust region  based on taking $E=1.6\times 10^6 \pm 25\%$ MPa, and the experimental data from the S4 system of \citet{OYE13} (symbols) with error bars smaller than the symbols according to \citet{OYE13}. The dotted line represents the standard lubrication-theory linear flow rate--pressure drop relation for a rigid rectangular channel, {\it i.e.}, $\Delta p = 12\mu \ell q/(h_0^3w)$.}
	\label{fig:expt_compare}
\end{figure}

To make a quantitative comparison between theory and experiment, figure~\ref{fig:expt_compare} shows a plot of the \emph{dimensional} flow rate--pressure drop relation \eqref{eq:q_of_p_dim} based on our asymptotic theory,  
the corresponding experimental measurements from the S4 system of \citet{OYE13} and a reference line corresponding to the lubrication-theory result for a solid rectangular channel. Clearly, there is good agreement between theory and experiment, especially with respect to the shape of the flow rate--pressure drop curve. In particular, the crossover from a linear (at low $q$) to a nonlinear (at high $q$) regime can be observed in the plot. 

The theoretical curve shown in figure~\ref{fig:expt_compare} systematically underpredicts the pressure drop by about $5\%$. We attribute this small discrepancy to two factors. First, the deformation of the S4 microchannel is not insignificant: as reported in table~1 of \citet{OYE13}, $\max_{x,z}|u(x,z)|/t \approx 0.43$. Such a maximum displacement, given the top wall's thickness, might be considered at the edge of applicability of the Kirchhoff--Love (linearly elastic) plate theory. Hence, the agreement might be improved if a different top-wall elasticity model, which accounts for the nontrivial wall thickness, were used. Second, the elasticity parameters of the PDMS microchannels are not well characterized by \citet{OYE13}. We have estimated a value of $E\approx 1.6$ MPa using their description of the experimental procedure, cross-referencing against the data on PDMS material properties tabulated by \citet{JMTT14} and treating the hydrostatic loading data \citep[figure~1({\it e},{\it f})]{OYE13} as a ``bulge test'' \citep{SN92}. However, there is inherent uncertainty in such an estimate, which is why we have added a confidence region (corresponding to a $\pm25\%$ variation in the Young's modulus $E$) around the theoretical curve in fig.~\ref{fig:expt_compare}. This uncertainty appears to capture most of the (small) discrepancy between theory and experiment. Therefore, it is clear that figure~\ref{fig:expt_compare} shows that a \emph{quantitative} prediction of the flow rate--pressure drop curve for a long shallow microchannel is possible {without fitting parameters}.

Finally, based on the discussion \S\ref{sec:previous_models}, it might be appropriate to think of the fitting parameter $\alpha$ of \citet{GEGJ06} as constant given by $\alpha = \tfrac{1}{60} (w/t)^3 (1-\nu^2)$. Then, for the S4 experiments of \citet{OYE13}, we find that fitting the data to \eqref{eq:q_p_gervais_dim} evaluated at $z=0$ yields $\alpha \approx 5.91$.\footnote{Here, we have used {\sc Mathematica}'s built-in nonlinear least-squares subroutine {\tt FindFit}. Note, however, that the best-fit value is sensitive to small errors introduced in digitizing the data from \citep{OYE13}.} Meanwhile, $\alpha = \tfrac{1}{60} (w/t)^3 (1-\nu^2) \approx 7.68$ using the values given in table~\ref{tb:params1}.

\section{Discussion and conclusions}
\label{sec:concl}

Using asymptotic techniques, we have derived a flow rate--pressure drop relation for low-Reynolds-number flow in a shallow deformable channel. Our result, which contains no fitting parameters, can be applied to the design and construction of soft microfluidic devices, where the flow and the elastic deformation of the domain are coupled. Specifically, we treated the case of bending-dominated deformation, which is governed by the plate equation \eqref{eq:KL_plate}. This restricts our analysis to microchannels for which the maximum deformation $u_\mathrm{max}:=\max_{x,z}|u(x,z)|$ is much smaller than the thickness of the top wall, which is, in turn, much smaller than the channel's width: $u_\mathrm{max} \ll t\ll w$. 

To make clear the limitations of our model, consider first the case when $t \gg w$, {\it i.e.}, the top wall is much thicker than the width of the channel, as is the case in the experiments of \citet{GEGJ06}. Then, the top wall behaves like an elastic half-space rather than a plate. An extension of the present work would require us to show a similar decoupling between the flow-wise and transverse deformation, as was done in \S\ref{sec:plate}, perhaps adapting the solution for a uniform load distributed over a finite width of an elastic half-space \citep[\S2.4]{J85}. Now, consider the case when $u_\mathrm{max}\gg t$, {\it i.e.}, the top wall's deformation is much larger than its thickness, as is the case in some of the experiments of \citet{OYE13}. Now, the top wall behaves like a membrane rather than a plate. Once again, an extension of the present work would be to revisit \S\ref{sec:plate} starting from a model of stretching-dominated deformation, perhaps adapting the approximate solution from \citep[Art.~101]{TWK59}. Each case would require a careful re-calculation of the flow rate--pressure drop relation and appropriate comparisons against applicable experiments. Unfortunately, at this time, few reliable measurements of the top-wall displacement of microchannels under hydrodynamic conditions have been published. 

Returning to our results for the bending-dominated case, in dimensional variables and keeping only the leading-order elasticity contribution in \eqref{eq:q_of_p_dim}, we can summarize the key result:
\begin{equation}
q \approx \frac{h_0^3 w\Delta p}{12\mu \ell} \left[1 + \frac{3}{160}\left(\frac{w}{t}\right)^3\left(\frac{w}{h_0}\right)\left(\frac{\Delta p}{E}\right) \right],
\label{eq:q-dp_simple}
\end{equation}
where $q$ is the flow rate, $\Delta p$ is the pressure drop, $w$ is the channel width, $h_0$ is the undeformed channel height, $\ell$ is the channel length, $t$ is the top wall's thickness, and $E$ is the Young's modulus of the material (assuming an incompressible material, \textit{i.e.}, a Poisson ratio $\nu = 1/2$). In particular, \eqref{eq:q-dp_simple} highlights the strong dependence on the plate geometry through the factor of $(w/t)^3$, which is assumed $\gg1$ in our plate-bending model. Note that the ratio outside the square brackets in \eqref{eq:q-dp_simple} represents the leading-order term (in $h_0/w\equiv\delta$) in the flow rate--pressure drop relation for a shallow rectangular channel. 

However, the boundary conditions in the transverse $x$-direction are not enforced in this approach. This defect comes from the shallowness assumption that allows us to neglect $\mathcal{O}(\delta^2)$ terms in the asymptotic expansion. A potential remedy is to relax the assumption that $\delta\ll 1$ and solve the full-2D leading-order flow problem using the {domain perturbation} technique \citep{VD75,L82}. An alternative approach, consistent with lubrication theory, is presented in Appendix~\ref{app:notshallow}. Finally, future work might include computing the remaining two components of the leading-order velocity field. This three-dimensional case requires further consideration because, as shown by \citet{LSS04}, if the cross-section is varying in the flow-wise direction, then the flow cannot be planar, so the continuity equation alone does not yield the remaining velocity component as in standard lubrication theory. It would also be of interest to compute higher-order perturbative corrections to the lubrication theory velocity profile given in \S\ref{sec:lubri} following, \textit{e.g.}, \citet{THFS17}. 



\section*{Acknowledgements}
This work was initiated while I.C.C.\ and V.C.\ were working with H.A.S.\ at Princeton University. At that time, I.C.C.\ was supported by the National Science Foundation (NSF) under grant No.\ DMS-1104047. H.A.S.\ thanks the NSF for support via grant No.\ CBET-1132835. 
We thank the anonymous referees for their incisive comments, questions and suggestions, which helped improve the manuscript. Specifically, one of the referee's remarks lead to the calculation in Appendix~\ref{app:notshallow}.

\appendix

\section{Explicit pressure and velocity expressions for a stiff top wall}
\label{sec:stiff_topwall}

\begin{figure}
	\centering
	\vspace{5mm}
	\includegraphics[width=0.5\textwidth]{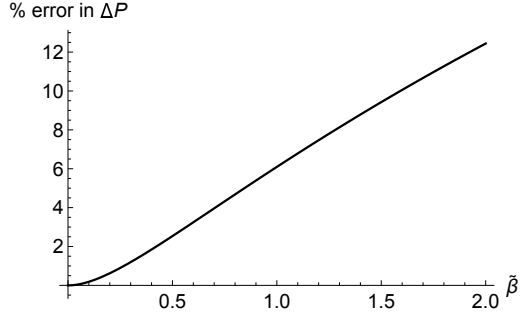}
	\caption{Percent error in the full pressure drop $\Delta P = P(0)$ as calculated via the approximate explicit relation \eqref{eq:p_of_z_small_a}, for a flow-rate controlled situation ($Q=1$), and compared to pressure drop from the ``full'' implicit relation \eqref{eq:q_of_dp2}, as a function of the compliance parameter $\tilde\beta$.}
	\label{fig:p_error}
\end{figure}

Note that the last two terms in the bracket on the right-hand side of \eqref{eq:q_of_p} have large denominators. Thus, let us neglect these two terms under the assumption that $\max_{0\le Z\le 1} \tilde\beta P(Z) \equiv \tilde\beta P(0)  = \mathcal{O}(1)$. This approximation reduces the flow rate--pressure drop relation to a quadratic equation that can be solved explicitly for $P$:
\begin {equation}
P(Z) = -\frac{10}{\tilde\beta} \left[ 1 \pm \sqrt{1 + \frac{12}{5}\tilde\beta Q (1-Z)} \,\right],
\label{eq:p_of_z_small_a}
\end{equation}
where we must pick the ``$-$'' sign above because our convention is that $P(1) = 0$. The approximate expression \eqref{eq:p_of_z_small_a} highlights the fact that $P$ is a \emph{nonlinear} function of $Z$ in a deformable channel, unlike the rigid-channel relation $P(Z)=12(1-Z)$, which is linear. The maximum error committed by \eqref{eq:p_of_z_small_a} is for $P(Z=0)$, which corresponds to the error in the full pressure drop $\Delta P = P(0)$. Figure~\ref{fig:p_error} shows the percent error in $\Delta P$ as predicted by \eqref{eq:p_of_z_small_a} compared to the result from \eqref{eq:q_of_dp2} in a flow-rate controlled situation ($Q=1$). It is evident that the error is modest ($\lesssim15\%$) for up to $\tilde{\beta}=2$.

Next, we expand \eqref{eq:p_of_z_small_a} into a Taylor series (for $\tilde\beta Q \ll1$) to highlight the first perturbative correction to the rigid-channel expression:
\begin{equation}
P(Z) = 12Q(1-Z)\left[1 - \frac{3}{5}\tilde{\beta}Q(1-Z) + \mathcal{O}\big(\tilde{\beta}^2Q^2\big)\right].
\label{eq:p_of_z_small_a2}
\end{equation}
Now, differentiating $P$ with respect to $Z$ from either \eqref{eq:p_of_z_small_a} or \eqref{eq:p_of_z_small_a2}, we find
\begin{equation}
-\frac{\rd P}{\rd Z} = 12 Q \left[ 1 + \frac{12}{5}\tilde\beta Q (1-Z) \right]^{-1/2}
= 12Q\left[1 - \frac{6}{5}\tilde{\beta}Q(1-Z) + \mathcal{O}\big(\tilde{\beta}^2Q^2\big)\right].
\label{eq:dpdz_small_a}
\end{equation}
Substituting $-\rd P/\rd Z$ from \eqref{eq:dpdz_small_a} into \eqref{eq:VZ00}, and using \eqref{eq:topwall_dimless} with $U$ given by \eqref{eq:U_of_XZ}, we obtain the velocity distribution
\begin{equation}
\begin{aligned}
V_Z(X,Y,Z) &= 6 Q \left[ 1 + \frac{12}{5}\tilde\beta Q (1-Z) \right]^{-1/2}\\ 
&\phantom{=}\times \left\{ 1 - 10 \left[1 - \sqrt{1 + \frac{12}{5}\tilde\beta Q (1-Z)} \,\right] \left( X + \tfrac{1}{2} \right)^2\left( X - \tfrac{1}{2} \right)^2 - Y\right\} Y.
\end{aligned}
\label{eq:vz_deformed}
\end{equation}
Now, we expand \eqref{eq:vz_deformed} into a Taylor series (for $\tilde\beta Q\ll1$) to highlight the first perturbative correction to the rigid-channel expression:
\begin{equation}
V_Z(X,Y,Z) = 6Q(1-Y)Y + \frac{9}{10}\tilde{\beta} Q^2 (80X^4 - 40X^2 + 8Y - 3)Y(1-Z) + \mathcal{O}(\tilde{\beta}^2Q^2).
\label{eq:vz_deformed_expanded}
\end{equation}
Clearly, the first correction to the velocity field, unlike the corresponding one to the pressure in \eqref{eq:p_of_z_small_a2}, introduces dependence upon \emph{both} of the coordinates (\textit{i.e.}, $X$ and $Z$) that are not present in the leading-order term. 

Equation \eqref{eq:vz_deformed} provides a tractable expression from which to compute the maximum velocity for a given $Z$ cross-section with $Q$ and $\tilde{\beta}$ as parameters:
\begin{equation}
V_{Z,\mathrm{max}}(Z;Q,\tilde{\beta}) := \max_{\substack{-1/2 < X < 1/2\\ 0 < Y < H(X,Z)}} V_Z(X,Y,Z).
\end{equation}
Performing this standard calculation using the approximate velocity expression \eqref{eq:vz_deformed} yields
\begin{equation}
V_{Z,\mathrm{max}}(Z;Q,\tilde{\beta}) \simeq 
\frac{3}{2}Q\left[1 + \frac{3}{10} \tilde{\beta} Q (1-Z)\right] \qquad (\tilde\beta Q\ll 1).
\label{eq:VZ_max}
\end{equation}
To the leading order in $\tilde\beta Q$, equation \eqref{eq:VZ_max} predicts that the dependence upon $Z$ vanishes at the downstream end of the microchannel, which is consistent with the fact that there is no deformation of the top wall there.

\section{Effect of the lateral side walls}
\label{app:notshallow}

The effect of the lateral side walls becomes important if the channel is not necessarily shallow, \textit{i.e.}, $\delta \equiv h_0/w \not\ll 1$. Returning to the fluid's governing equations \eqref{eq:divV_dimless}--\eqref{eq:stokes_dimless3}, we can retain terms involving $\delta$, while still only considering the equations at the leading order in  $\epsilon \equiv h_0/\ell \ll 1$. In other words, the Stokes equations \eqref{eq:stokes_dimless3} reduce to
\refstepcounter{equation}
$$
\delta^2 \frac{\partial^2 V_Z}{\partial X^2} + \frac{\partial^2 V_{Z}}{\partial Y^2} = \frac{\partial P}{\partial Z},\qquad
0 = \frac{\partial P}{\partial Y},\qquad  0 = \frac{\partial P}{\partial X},
  \eqno{(\theequation{\mathit{a},\mathit{b},\mathit{c}})}
\label{eq:leading_order_2D}
$$
and the continuity equation \eqref{eq:divV_dimless} remains unchanged. Clearly, calculating the flow rate--pressure drop relation for channels that are not necessarily shallow requires solving a two-dimensional (2D) problem since the flow profile depends on both the $X$ and $Y$ cross-sectional coordinates. Unfortunately, the problem is posed on a non-rectangular domain: $\{(X,Y) \;| -1/2 \le X \le +1/2,\; 0\le Y \le H(X,Z) \}$, where $H(X,Z) \equiv 1 + \beta U(X,Z)$ with the displacement $U(X,Z)$ computed self-consistently through \eqref{eq:beam2}, which remains unchanged even if $\delta=\mathcal{O}(1)$. The no-slip condition provides boundary conditions at the walls: $V_Z(\pm 1/2,Y) = V_Z(X,0) = V_Z\big(X,H(X,Z)\big) = 0$.

Although the difficulty of solving a 2D problem on a non-rectangular domain can be overcome by the {domain perturbation} technique \citep{VD75,L82}, this approach would require that we expand the channel's shape in a power series in $\beta\ll1$ as well ({\it i.e.} a \emph{nearly rectangular} domain). However, $\beta \not\ll 1$ for this problem, as discussed above, for any $\delta$ including $\delta = \mathcal{O}(1)$. Instead, we proceed by a lubrication argument in which $\beta$ can be large (how large will be specified below) but $\delta \ll 1$.

To this end, first we solve \eqref{eq:leading_order_2D} on the rectangular domain $(X,Y) \in [-1/2,+1/2]\times[0,1]$. This problem has a known solution, which can be found by separation of variables \citep[{\it e.g.},][\S3.4.6]{B08}:
\begin{multline}
V_Z(X,Y)  = -\frac{\rd P}{\rd Z} \Bigg\{ \frac{1}{2}(1-Y)Y \\ - \frac{4}{\pi^3} \sum_{n=1}^\infty \frac{1}{(2n-1)^3}\frac{\cosh[(2n-1)\pi X/\delta]}{\cosh[(2n-1)\pi/(2\delta)]} \sin[(2n-1)\pi Y] \Bigg\},
\label{eq:VZ_2D}
\end{multline}
hence,
\refstepcounter{equation}
$$
  Q = -\frac{1}{12}\frac{\rd P}{\rd Z} \left[ 1 - \kappa(\delta)\right],\qquad \kappa(\delta) := \sum_{n=1}^\infty \frac{1}{(2n-1)^5} \frac{192}{\pi^5} \delta \tanh\left[\frac{(2n-1)\pi}{2\delta}\right]
  \eqno{(\mathrm{\theequation}{\mathit{a},\mathit{b})}}
\label{eq:Q_2D}
$$
for a channel of fixed rectangular cross-section with aspect ratio $\delta=h_0/w$. Furthermore, for future reference, we note the approximation $\kappa(\delta) \approx \kappa_0\delta$ as $\delta \to 0^+$, where $\kappa_0 = (192/\pi^5)(31/32)\zeta(5) \approx 0.630$, and $\zeta(\xi) \equiv \sum_{n=1}^\infty n^{-\xi}$ is the Riemann zeta function \cite[\S23.2]{AS72}.

For $\delta\ll1$, the terms in the summation over $n$ in \eqref{eq:VZ_2D} lead to small corrections that are localized in a boundary layer near the side walls $X=\pm1/2$. [Of course, from \eqref{eq:stokes_dimless3_Z}, it is evident that a boundary-layer calculation can be done by the rescaling $X\mapsto \delta X$, which keeps the first term on the left-hand side of \eqref{eq:stokes_dimless3_Z}, to arrive precisely at \eqref{eq:VZ_2D} for a rigid channel.] The net effect, which is captured in \eqref{eq:Q_2D}, is that the flow rate is reduced by $\kappa(\delta)$ due to friction at the $X=\pm1/2$ sidewalls. Moreover, since the top wall is assumed to be clamped and, thus, $H(X,Z) \to 1$ as $X\to\pm1/2$, then the latter observation is true even for a \emph{shape-deformed} channel in which $Y\in[0,H(X,Z)]$. 

Thus, within the lubrication approximation, if we wish to estimate the leading-order contribution of drag due to the sidewalls at $X=\pm1/2$ even in the case of a {deformable} cross-section, then we can employ \eqref{eq:Q_2D}, provided that we ``rescale'' it to the domain $\{(X,Y) \;| -1/2 \le X \le +1/2,\; 0\le Y \le H(X,Z) \}$:
\begin{multline}
V_Z(X,Y,Z)  = -\frac{\rd P}{\rd Z} \Bigg\{ \frac{1}{2}[H(X,Z)-Y]Y \\ - \frac{4}{\pi^3} \sum_{n=1}^\infty \frac{1}{(2n-1)^3}\frac{\cosh[(2n-1)\pi X/\delta]}{\cosh[(2n-1)\pi/(2\delta)]} \sin\left[(2n-1)\pi \frac{Y}{H(X,Z)}\right] \Bigg\}.
\label{eq:VZ_2D_deform}
\end{multline}

Why should \eqref{eq:VZ_2D_deform} apply? Obviously, in the limit as $\delta\to0^+$, we obtain the result in \S\ref{sec:lubri}. For $\delta\ne0$, we must now consider the summation term. If $\delta \ll 1$, then the summation term in \eqref{eq:VZ_2D_deform} is localized in boundary layers near $X=\pm1/2$, \emph{independently} of $Y$. It is easy to verify that \eqref{eq:VZ_2D_deform} satisfies all the boundary conditions, namely $V_Z(\pm1/2,Y)=V_Z(X,0)=V_Z\big(X,H(X,Z)\big)=0$. However, the velocity profile \eqref{eq:VZ_2D_deform} does \emph{not} satisfy the governing equation \eqref{eq:leading_order_2D}. The error committed is $\mathcal{O}(\beta\delta)$. Therefore, if $\delta\ll1$ ($\Rightarrow 1/\delta\gg1$), then \eqref{eq:VZ_2D_deform} is asymptotically valid for $\beta$ as large as $\beta = \mathcal{O}(1/\delta)$; in other words, \eqref{eq:VZ_2D_deform} applies for $\beta \ll 1/\delta$, or $u_\mathrm{c} \ll w/\delta^2$ (\textit{i.e.}, the deformation does not necessarily have to be small).

Substituting \eqref{eq:VZ_2D_deform} into \eqref{eq:flow_rate}, we find the flow rate:
\begin{multline}
Q = -\frac{1}{12}\frac{\rd P}{\rd Z}\Bigg[1 + \frac{\tilde\beta}{10} P(Z) + \frac{\tilde\beta^2}{210} P(Z)^2 + \frac{\tilde\beta^3}{12,012} P(Z)^3 \\
- \frac{96}{\pi^4} \sum_{n=1}^\infty \frac{1}{(2n-1)^4} \int_{-1/2}^{+1/2} \frac{\cosh[(2n-1)\pi X/\delta]}{\cosh[(2n-1)\pi/(2\delta)]} H(X,Z) \,\rd X \Bigg].
\label{eq:Q_dPdZ_appendix}
\end{multline}
The last integral can be evaluated exactly based on \eqref{eq:U_of_XZ}, and it gives the correction to \eqref{eq:q_of_dpdz} due to the lateral side walls at $X=\pm1/2$. For the purposes of finding the leading-order correction (in $\delta\ll1$), it suffices to note that
\begin{equation}
\int_{-1/2}^{+1/2} \frac{\cosh[(2n-1)\pi X/\delta]}{\cosh[(2n-1)\pi/(2\delta)]} H(X,Z) \,\rd X  = \frac{2}{(2n-1)\pi} \delta \tanh\left[\frac{(2n-1)\pi}{2\delta}\right] + \mathcal{O}(\delta^3).
\end{equation}
Thus, \eqref{eq:q_of_p} corrected to the leading-order contribution due to drag at the sidewalls at $X=\pm1/2$ in the case of a \emph{deformable} cross-section, is found by integrating \eqref{eq:Q_dPdZ_appendix} with respect to $Z$ subject to $P(1)=0$ and neglecting terms of $\mathcal{O}(\delta^3)$:
\begin{equation}
Q = \frac{P(Z)}{12(1-Z)}\left[ 1 + \frac{\tilde\beta}{20}P(Z) + \frac{\tilde\beta^2}{630} P(Z)^2 + \frac{\tilde\beta^3}{48,048} P(Z)^3 - \kappa_0\delta \right].
\label{eq:q_of_p_corr_lubri}
\end{equation}

The approach we have outlined in this appendix should be contrasted to how the effect of sidewalls has been treated by \citet{CTS12} and the literature derived thereof \citep[{\it e.g.},][]{RS16,RDC17}.  The approach based on \citep{CTS12} is an extension of the \citet{GEGJ06} model to channels that are not necessarily shallow, meaning it does not account for the details of the deformation profile in the cross-section. \citet{CTS12} include the factor $[1-\kappa(\delta)]\approx[1-\kappa_0 \delta]$ from \eqref{eq:Q_2D} in the integral defining the flow rate, and then account for the variable channel height by replacing it with $[1-\kappa_0 \delta H(X,Z)]$ . This translates into our analysis as multiplying the integrand of \eqref{eq:flow_rate2} by $[1-\kappa_0\delta H(X,Z)]$. Using the expression for the channel shape given in \eqref{eq:topwall_dimless}, we  obtain
\begin{equation}
Q  = -\frac{1}{12} \frac{\rd P}{\rd Z} \left\{  
\int_{-1/{2}}^{+1/{2}} [1+\beta U(X,Z)]^3\big(1-\kappa_0\delta [1+\beta U(X,Z)]\big) \,\rd X \right\}.
\label{eq:flow_rate_int_h_corr}
\end{equation}
Then, substituting the expression for $U(X,Z)$ from \eqref{eq:U_of_XZ} into \eqref{eq:flow_rate_int_h_corr} and carrying out the integrations over $X$ and $Z$, we find
\begin{multline}
Q = \frac{P(Z)}{12(1-Z)}\Bigg\{ 1 + \frac{\tilde\beta}{20} P(Z) + \frac{\tilde\beta^2}{630} P(Z)^2 + \frac{\tilde\beta^3}{48,048} P(Z)^3 \\
- \kappa_0\delta \left[1 + \frac{\tilde\beta}{15} P(Z) + \frac{\tilde\beta^2}{315} P(Z)^2 + \frac{\tilde\beta^3}{12,012} P(Z)^3 + \frac{\tilde\beta^4}{1,093,950} P(Z)^4 \right] \Bigg\}.
\label{eq:q_of_p_corr}
\end{multline}
Equation \eqref{eq:q_of_p_corr} could be compared to equation (10) of \citet{CTS12}. However, this  approximation has the effect of \emph{overestimating} the volumetric flow rate loss due drag at the sidewalls, as the factor $[1-\kappa_0\delta H(X,Z)]$ now modifies the flux even at the center of the channel, which in turn introduces the four terms depending on $P(Z)$ in the brackets in \eqref{eq:q_of_p_corr}.

In summary, we have shown via a perturbation expansion in this Appendix that, to leading order in $\delta\ll 1$, equation \eqref{eq:q_of_p_corr_lubri}, rather than \eqref{eq:q_of_p_corr}, contains the asymptotic correction to the flow rate.

\bibliography{deformable_channel_jfm}
\bibliographystyle{jfm}
\end{document}